\documentclass[runningheads]{llncs}
\usepackage[T1]{fontenc}
\usepackage{mathtools}
\usepackage{caption}
\usepackage{subcaption}
\usepackage{graphicx}
\usepackage{amsmath}
\usepackage{amssymb}
\usepackage{amsfonts}
\usepackage{algorithm, algcompatible}
\usepackage{comment}
\begin{document}
%dynamic preference for 
\title{Dynamic Preference Multi-Objective Reinforcement Learning for Internet Network Management}
\titlerunning{Dynamic Preference Multi-Objective Reinforcement Learning}
% If the paper title is too long for the running head, you can set
% an abbreviated paper title here
%
\author{DongNyeong Heo\orcidID{0000-0002-8765-6744} \and
Daniela Noemi Rim\orcidID{0000-0002-4527-8028} \and
Heeyoul Choi\orcidID{0000-0002-0855-8725}}
\authorrunning{D. Heo et al.}
% First names are abbreviated in the running head.
% If there are more than two authors, 'et al.' is used.
%
\institute{Handong Global University, Pohang, South Korea\\
\email{\{dnheo, danielarim, hchoi\}@handong.ac.kr}}
\maketitle              % typeset the header of the contribution
\begin{abstract}
%The abstract should briefly summarize the contents of the paper in
%150--250 words.

An internet network service provider manages its network with multiple objectives, such as high quality of service (QoS) and minimum computing resource usage. To achieve these objectives, a reinforcement learning-based (RL) algorithm has been proposed to train its network management agent. Usually, their algorithms optimize their agents with respect to a single static reward formulation consisting of multiple objectives with fixed importance factors, which we call \textit{preferences}. However, in practice, the preference could vary according to network status, external concerns and so on. For example, when a server shuts down and it can cause other servers' traffic overloads leading to additional shutdowns, it is plausible to reduce the preference of QoS while increasing the preference of minimum computing resource usages. In this paper, we propose new RL-based network management agents that can select actions based on both states and preferences. With our proposed approach, we expect a single agent to generalize on various states and preferences. Furthermore, we propose a numerical method that can estimate the distribution of preference that is advantageous for unbiased training. Our experiment results show that the RL agents trained based on our proposed approach significantly generalize better with various preferences than the previous RL approaches, which assume static preference during training. Moreover, we demonstrate several analyses that show the advantages of our numerical estimation method. 
%In reinforcement learning (RL), addressing multi-term reward functions presents a challenge when each term has a different impact depending on the context where users want to customize the reward function accordingly. 
%In such scenarios, conventional approaches with fixed parameters fall short, as the agent cannot dynamically adapt to the user's input.
%In this study, we consider user's input as additional parameters, and the RL agent's primary objective is to maximize the expected return in accordance to these parameters. To address this user-adjustable RL (UARL) scenario, we propose incorporating the parameters into the RL agent's policy and value networks during training, treating it as part of the state representation. Consequently, our RL agent gains the ability to flexibly adjust its policy based on the user's input parameters.
%Experiment results in the context of network management tasks reveal that our proposed method is suitable to address the user-adjustable RL scenario without re-training compared to conventional RL methods.
%The qualitative analysis shows that our agent works flexibly in virtual network management scenarios where the reward function can change.

\end{abstract}
%
%
%page limit of 12-15 INCLUDING references
\section{Introduction}

A typical network management system comprises several sub-modules, such as service function chaining (SFC) \cite{heo2020graph}, network function (e.g., firewall, proxy, etc.) deployment, anomaly detection \cite{lee2021sequential}, auto-scaling (AS) \cite{seo2022updating}, and power management (PM) \cite{abbas2022autonomous}. 
Based on the emergence of huge-scale internet networks, network service management modules have integrated machine learning algorithms, such as reinforcement learning (RL) \cite{heo2020reinforcement,seo2022updating}, to optimally provide satisfying quality of service (QoS) to users and manage to reduce computing resource usages. To optimize the RL agent with respect to a scalar reward value, the above objectives measured in various ways (e.g., delay time of the service, amount of CPU usage, power consumption, etc.) are usually summed up with importance factors (which we call \textit{preferences}) as coefficients. Previous RL approaches optimized their agents with respect to rewards that have static formulations with fixed preferences.
%Such orchestration is modular and subjected to the agreed service level between the service provider and the clients (Service Level Agreement or SLA), and an efficient and reliable quality of service (QoS - quantified by amount of transferred data, time consumed for tasks, etc.) Specific modules monitor, control, deploy, and remove services based on SLA and QoS metrics. For example, the Service Function Chaining (SFC) module determines efficient paths for processing requested functions \cite{heo2020graph}. The auto-scaling module then adjusts resource allocation based on current demand \cite{lee2020deep}, while the Power Management module monitors energy consumption \cite{abbas2022autonomous}. 

%The objective of a sub-module is usually designed by multiple objectives,
%Each objective of a network sub-module is usually designed with multiple terms, such as high QoS and minimum computing resource usages. Previous RL applications optimized their agents with respect to static reward formulations, %that are composed of multiple objectives with coefficients that control each objective's importance (which we call \textit{preference}). 
%that is, each term within a reward had coefficients controlling each objective's importance (which we call \textit{preference}). 
However, in a realistic network service management scenario, the preference of an objective often varies based on the situation. For example, suppose a sudden shutdown of a computing server occurs. In that case, the preference for reducing computing resource usage increases to prevent overloads of other computing servers and chains of subsequent shutdowns. In this scenario, the previous RL approaches cannot consistently guarantee stable actions. Alternatively, multiple agents can be trained on multiple preferences that are manually set for various situations and can be utilized in each situation. However, this idea requires a lot of computing costs for the multiple training, and it would be unclear which RL agent trained with a certain preference will optimally behave in an unexpected situation. Therefore, there is a need for an RL algorithm that can train a single agent to behave generally well with various preferences.
%In a realistic scenario, the network modules operate concurrently with specialized (but not necessarily mutually exclusive) multi-term objective functions related to resource usage. The importance of each term within a module's objective should vary automatically based on specific service constraints %or according to an external agent's preference. 

%There are several approaches in RL to deal with multi-objectives, e.g. by treating the system as a single objective task via scalarizing reward terms or learning a set of optimal policies that can span the space of preferences \cite{van2013scalarized,natarajan2005dynamic,mossalam2016multi,zuluaga2016pal,reymond2019pareto}. However, the first one requires knowing beforehand the distribution of preference w.r.t. objectives, and the latter lack scalability in high-dimensional environments \cite{buet2023robust}. 
To address the dynamic changes of preference during training and testing, the robust dynamic preference multi-objective RL (RDP MORL) algorithm was introduced \cite{buet2023robust}. RDP MORL learns a single parameterized policy encompassing the dynamic of preference and action by introducing a surrogate space containing both the task states and preferences. During training, their algorithm samples a value of preference from a known or estimated distribution and trains the agent to take an optimal action in accordance with the sampled preference. They applied this framework to the Q-learning algorithm \cite{Sutton1998}, and showed improved generalizability of the RL agent toward various preferences. %This framework can be applied to both Q-learning and actor-critic methods, with the authors focusing on the former while we focus on the latter. Consequently, the agent can generalize to various preferences. 

%In this paper, we leverage the application of RDP MORL framework to train the sub-modules of network service management, such as AS and PM. 
In this paper, we propose a similar training framework to RDP MORL for the advanced actor-critic algorithm, proximal policy optimization (PPO) \cite{ppo2017}, and apply it to real-world problems, the sub-modules of network service management, such as AS and PM.
Specifically, we consider the preference of each network management objective as a variable and input those values to the policy and value networks. Therefore, we make our network management RL agents take actions dynamically according to various preferences. In addition, to ensure our agent can generalize without bias across preferences, we propose the numerical method that can estimate the preference distribution leading to better generalizability across preferences. 

The experiment results show that our proposed method performs well in static and dynamic preference testing scenarios, while the conventional RL approach can perform properly only in a static preference scenario that the agent was trained on. In both quantitative and qualitative analyses, the agent trained with our proposed method shows significant adaptability toward different preferences in terms of reward and actual outcomes in the network service management environment. Additionally, the experimental comparison between the two usages of our proposed numerically estimated preference distribution and an arbitrary uniform distribution demonstrates the advantage of our numerical method.

\section{Background}

\subsection{Multi-Objective Reinforcement Learning}
\label{subsec:rdp_morl}

%the reader doesn't know concept of preference, but they know MORL
Multi-objective RL (MORL) refers to scenarios where the total reward function comprises $b$ multiple sub-reward terms, each addressing different aspects of the task, to learn policies that simultaneously optimize several criteria \cite{yang2019generalized}. There are different applications of MORL depending on the importance (or \textit{preference}) of each component in the total reward. For example, the importance can be controlled and shaped linearly \cite{van2013scalarized}, and even vary dynamically 
%in a dynamic preference scenarioas the previous work,
as in RDP MORL \cite{buet2023robust}
%, concerned. 
Our work focuses on the latter case.

RDP MORL uses a surrogate state space $\hat{\mathcal{S}}$ containing both state $\mathcal{S}$ and preference $\Omega$ spaces 
%(with a joint transition function $\mathcal{P}_S \otimes \mathcal{P}_\omega$) 
allowing the application of the conventional RL algorithms to this dynamic preference scenario. Furthermore, it jointly inputs the preference $\omega_t \in \Omega$ with the state $s_t \in \mathcal{S}$ to the action value network, reflecting a policy preference at each time step. The initial value of $\omega_t$ at each episode is sampled from a certain distribution $p(\omega)$ that is given. 
%or estimated by an exploration model.

%RDP MORL was applied to the Q-learning algorithm \cite{buet2023robust} in each episode by sampling a preference from a certain distribution $G_\omega$. Given a state $s_t$, a sample $(s_t, \omega_t)$ is derived from the surrogate space and given as input to the $Q$ function from which an action $a_t$ is derived with probability $\epsilon$. That is, the agent takes an action influenced by the preference $\omega_t$ at a time $t$. The training regime subsequently follows a Q-learning algorithm.

%After executing the action, we observe a reward $\mathbf{r}_t$, stat $s_{t+1}$ and sample a preference $w_{t+1}$ from a transition distribution. Transitions are stored, adn a mini batch of transitions is sampled. for each sampled transition with its preference, sample a preference and future preference, and 

The network service setting has MORL scenarios, and we propose a MORL method for the network service management.
%During our research on MORL in the network service setting, we independently conducted a similar study to RDP MORL \cite{buet2023robust}.
Unlike RDP MORL, we apply our algorithm to advanced actor-critic methods like PPO, rather than Q-learning used in RDP MORL. Additionally, we examine how the choice of preference distribution, $p(\omega)$, affects the dynamic preference RL agent and propose a numerical estimation method for the preference distribution. Also, we use $b-1$ intrinsic dimensions for $\Omega$ with $b$ sub-reward terms, fixing the first preference to 1. This relative preference allows us to express the other $b-1$ objectives relative to the first sub-reward. This approach simplifies exploration compared to using $b$ dimensions.

%During our research on multi-objective RL in the network service environment, we independently conducted a similar study to RDP MORL \cite{buet2023robust}. Slightly different from RDP MORL work, we employ $b-1$ intrinsic dimensions for $\Omega$ for $b$ sub-reward terms by fixing the first preference to 1. Because the preference is relative, we can represent the preferences of other $b-1$ objectives with respect to the preference of the first sub-reward, which is 1. In addition, it is advantageous for simpler exploration than the one in $b$-dimensional space. Also, we apply this algorithm to the advanced actor-critic regime, such as PPO, instead of Q-learning that the previous work targeted. Finally, we study how much the choice of preference distribution, $p(\omega)$, can affect the dynamic preference RL agent. 

\begin{figure}[h]
    \centering 
    \includegraphics[width=0.7\linewidth]{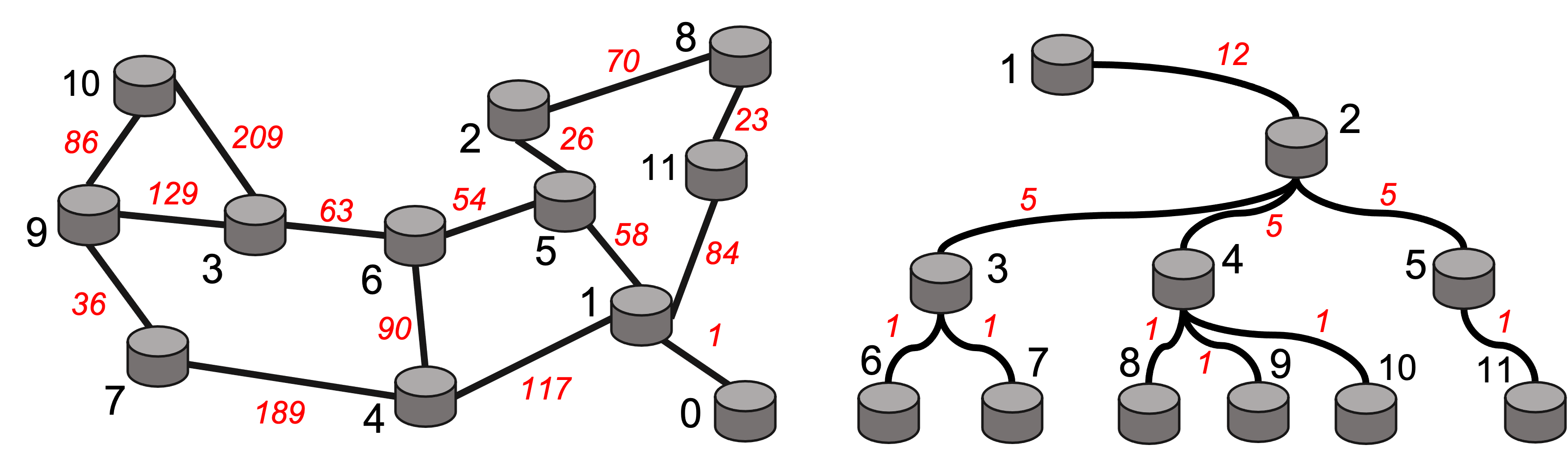}
    \hbox{\small (a) Internet2 Topology \hspace{0.7in} (b) MEC Topology}
    \caption{Network topologies that we used in this work. The black numbers indicate the index of the nodes, and the red numbers indicate the latency of the edge.}
    \label{fig:network_topologies}
\end{figure}

\subsection{Network Service Management}

Based on the technology of network function virtualization \cite{mijumbi2015network}, hardware-based network functions (e.g., firewall, proxy) are virtualized and deployed as a virtual machine instance in a computer server, which is called a virtualized network function (VNF) instance. The service provider builds a network by connecting computer servers and deploys different types of VNFs. Fig. \ref{fig:network_topologies} shows examples of networks that we used in this work, Internet2 and mobile edge computing (MEC) topologies. In the network, network traffic traverses between connected servers being operated by the deployed VNFs. The user requests a network service in the format of four elements: (1) source/destination servers, (2) bandwidth of traffic, (3) the service type that is a specific order of VNF types, and (4) requested service-level agreement (e.g., an amount of delay time of traffic traversing from the source to the destination servers). The service provider manages the network to ensure good QoS, that is successfully carrying out the service request, with minimal capital expenditure (CAPEX) and operating expenditure (OPEX), which are usually expensive in proportion to the amount of resource usage (e.g., the number of servers, the amount of deployed VNF instances, and electric power consumption). 

The entire management system usually consists of sub-modules, such as service function chaining (SFC) \cite{heo2020graph,heo2020reinforcement}, anomaly detection \cite{lee2021sequential}, VNF migration \cite{rim4633582anomaly}, VNF deployment, auto-scaling (AS) \cite{seo2022updating}, and electric power management (PM). According to the emergence of huge-scale internet networks, the sub-modules have been automated based on machine learning algorithms, such as RL. The reward function used to train the RL-based sub-module agent is usually formulated with multiple sub-reward terms, such as QoS and resource usage. As we depicted in the introduction, the importance (preference) of the sub-rewards could be dynamically changed based on the service provider's decision or external factors. In this paper, we propose our approach to train an RL agent that generalizes well to various preferences, especially for AS and PM sub-modules.

Before introducing the previous approaches of AS and PM sub-modules, we set mathematical notations we used throughout this paper. The internet network can be represented by a graph topology consisting of nodes (servers) and edges (connections between servers). Let $\mathcal{N}=\{n_1, \cdots, n_{|\mathcal{N}|}\}$ be the set of nodes in the topology. The edge set is composed of existing edges that are connections between connected nodes. Each edge has a delay time for traversing network traffic, we define the delay function $e(n_i,n_j)$ that maps the two pair of connected nodes, $n_i$ and $n_j$, to its edge's delay time. Let $\mathcal{F}=\{firewall, IDS, \cdots \}$ be the set of VNF types. Let $\mathcal{Q}=\{q_1, q_2, \cdots \}$ be the set of received service requests from users. According to a service request, a network traffic traverses from the source node to the destination node with intermediate nodes. This path of traffic is called SFC path and is represented by a sequence of nodes, $p_q=\{n_1^{q},n_2^{q}, \cdots \}$ where $n_i^{q}$ means the $i$-th node in the path of the request $q$'s network traffic. The traffic is processed by the VNF instances that are deployed in some nodes in the path. Then, the total delay of this SFC path is computed by summing up every edge's delays within the path, $\zeta(p_q)=\sum_{i=2}^{|p_q|}e(n_{i-1}^{q},n_{i}^{q})$. The requested service-level agreement of each service request is arbitrarily defined by the user or a manual, we represent it with $\zeta_{SLA}$.

\subsubsection{Auto-scaling Task} \label{subsubsec:autoscaling_task}
The AS sub-module scales the deployed VNF instances whenever a new service request is received. For example, if the demand of firewall VNF type is increased by new service requests, then it increases the capacity of existing firewall VNF instance or deploy a new firewall VNF instance at the optimal node. There are three types of scaling actions, deletion (scale-in), preservation (scale-keep), and addition (scale-out). An RL-based AS approach was proposed in \cite{seo2022updating}, and showed its feasibility in the network simulation dataset. In this paper, we use the main RL training framework of the previous work \cite{seo2022updating}. Therefore, we will briefly review the previous work's approach in this section.

The state, action, and reward for the AS RL approach's MDP are defined as follows:
\begin{itemize}
    \item State $S=\{S_1,S_2, \cdots \}$ where $S_q=(M,X_q)$ : Adjacency matrix $M \in \mathbb{R}^{|\mathcal{N}| \times |\mathcal{N}|}$, and annotation matrix $X_q \in \mathbb{N}^{|\mathcal{N}| \times (|\mathcal{F}|+2)}$ for every service requests.
    \item Action $A \in \{-1,0,1\}^{|\mathcal{N}\times\mathcal{F}|}$: Scale-in/-keep/-out for every node and VNF types.
    \item Reward $R$: (1) QoS represented by the negative average total delay of the SFC path divided by SLA, and (2) resource usages represented by the negative number of deployed VNF instances as follows,
    \begin{equation}\label{eq:auto_scaling_reward}
    R = -\frac{1}{|\mathcal{Q}|}\sum_{q \in \mathcal{Q}}{\frac{\zeta({p_q})}{\zeta_{SLA}}}-\alpha\sum_{f \in \mathcal{F}}{N(f)}.
    \end{equation}
\end{itemize}
The adjacency matrix represents the features of edges, each element is set by the inverse delay of the edge, $1/e(n_i,n_j)$. The annotation matrix represents the features of nodes that are related to each VNF type and source/destination node. The first to $|\mathcal{F}|$-th elements of a node feature are integers that are the number of deployed VNF instances of each type. The next two elements are binary numbers that indicate the source/destination nodes. To calculate the delay of the SFC path, $\zeta(p_q)$, we need a SFC path, $p_q$, generating algorithm. In the previous work, the external pre-trained graph neural network-based SFC model \cite{heo2020graph} was utilized. $N(f)$ is the current number of deployed VNF instances of the VNF type $f$ in the entire network topology. $\alpha$ controls the preference of the resource usage term, which is a pre-defined constant hyperparameter in the previous work. On the other hand, in our approach, we consider $\alpha$ as a variable.

The previous work used proximal policy optimization (PPO) \cite{ppo2017} algorithm for the training. For the policy (actor) model, the gated graph neural network-based (GGNN, \cite{li2015gated}) encoder-decoder model was utilized. The GGNN encoder receives the state, $S_q=(M, X_q)$, and outputs the hidden representation matrix for the service request $q$. The encoder repeats this process across every service request and aggregates the encoded hidden representations to output one hidden representation matrix. Then, the hidden representation is repeated and each representation is concatenated with embedding vectors of each VNF type. Finally, the feed-forward neural network-based (FF) decoder computes the probabilities of actions $A$ of every node and VNF type. These processes are formulated as follows:
\begin{align}
    &H=\frac{1}{|\mathcal{Q}|}\sum_{q \in \mathcal{Q}}GGNN(M,X_q) \in \mathbb{R}^{|\mathcal{N}| \times d}, \label{eq:ggnn_encoder} \\
    &Z=\left[ Re \left( H,|\mathcal{F}| \right)^T, Re \left( E_{\mathcal{F}},|\mathcal{N}| \right) \right] \in \mathbb{R}^{|\mathcal{N}| \times |\mathcal{F}| \times 2d}, \label{eq:concatenation} \\
    &\pi_{\theta}(A|S)=Softmax(FF(Z)), \label{eq:decoder}
\end{align}
where $d$ is the hidden dimension. $E_{\mathcal{F}} \in \mathbb{R}^{|\mathcal{F}| \times d}$ are embedding vectors of each VNF type \cite{mikolov2013efficient}. $[\cdot,\cdot]$ is the concatenation function. $Re(\cdot,k)$ is a function that repeats input tensor $k$ times.
Additionally, the value network $V_{\phi}(S)$ was designed to share the encoder with the actor, but it has a separated $FF$ layer that estimates the value function. 

\subsubsection{Power Management Task} \label{subsubsec:power_management_task}
The PM sub-module controls the total network system's power consumption. Usually, the power consumption is proportionally correlated with each computer server's CPU load levels. In the network management setting, these levels can be estimated with a public CPU specification, as done by \cite{abbas2022autonomous}. In the work \cite{abbas2022autonomous}, they automatized (using RL) the network system by monitoring the overall CPU energy consumption. Specifically, the RL agent learned to force idle devices into power-save mode while maintaining QoS. However, its agent is trained in a non-end-to-end fashion which may complicate the training process. Therefore, in this paper, we propose a new end-to-end training algorithm of the PM sub-module based on the AS framework (Section \ref{subsubsec:autoscaling_task}). Our approach borrows the way of obtaining the data for energy consumption from \cite{abbas2022autonomous}, and we include this measure in our reward formulation.

\section{Proposed Method}
\label{sec:proposed_method}
We consider the real-world scenario of network management where the preferences (coefficients in the reward formulation) of each sub-module can vary during training and testing. Therefore, our training objective is to train an agent that maximizes the expectation of reward with respect to state, action, and preference. The objective formulation below is an example of the objective of the AS task in the dynamic preference scenario:
\begin{equation}\label{eq:objective_of_our_scenario}
    \max \mathbb{E}_{\alpha \sim p(\alpha)}\mathbb{E}_{S \sim p(S)}\mathbb{E}_{A \sim \pi_{\theta}(A|S)}\left[R(\alpha)|S,A\right]. 
\end{equation}
In order to enforce the agent to achieve this goal, we train our agent with various preferences in a similar way the RDP MORL was trained. %based on an application of the RDP MORL algorithm. 
We call our AS and PM sub-modules {\em dynamic preference AS (DP-AS)} and {\em dynamic preference PM (DP-PM)}, respectively, that are targeted to dynamic preference scenarios. 

In this section, we describe how to train the DP-AS sub-module in Section \ref{subsec:dynamic_preference_auto_scaling_agent}. In Section \ref{subsec:power_management_agent}, we propose the new framework of PM sub-module based on the previous AS sub-module \cite{seo2022updating}, and we describe how to train the DP-PM sub-module. The task of DP-PM could be understood as generalizing the 1-dimensional preference space of DP-AS to the 2-dimensional preference space. Lastly, we discuss the importance of choosing the preference distribution, $p(\alpha)$, and propose a numerical method to select a proper preference distribution to ensure generalizability over a wide range of preferences in Section \ref{subsec:preference_distribution}. 

\subsection{Dynamic Preference Auto-Scaling Agent}
\label{subsec:dynamic_preference_auto_scaling_agent}

\subsubsection{Network Modification}
%As the original RDP MORL approach proposed, w
We interpret the preference as the additional element of state, and input the preference to the policy and value networks. The DP-AS agent has the preference $\alpha$ in its reward formulation as in Eq. \ref{eq:auto_scaling_reward}. We simply concatenate the preference value to every hidden representation vector at the aggregation process of the encoder (Eq. \ref{eq:concatenation}) as follows:
\begin{align}
    \hat{Z} &= [Z,Re(Re(\hat{\alpha},|\mathcal{F}|),|\mathcal{N}|)] \in \mathbb{R}^{|\mathcal{N}| \times |\mathcal{F}| \times (2d+1)}, \label{eq:concatenation_alpha}
\end{align}
where $\hat{\alpha}$ is normalized value of input $\alpha$. This normalization step is necessary to match the scale with the scales of other input elements, $M$ and $X$. Specifically, we divide the $\alpha$ value with the expectation of its given distribution, $\hat{\alpha}=\frac{\alpha}{\mathbb{E}\left[\alpha\right]}$, so that the expectation of this normalized value is 1. Finally, the policy and value networks compute action probabilities and the value function given $\hat{Z}$ instead of $Z$.

\begin{algorithm}
    \caption{DP-AS Agent Training Algorithm}
    \label{alg:training_algorithm}
    \begin{algorithmic}[1]
        \REQUIRE  Preference distribution $p(\alpha)$, simulation dataset $\mathcal{D}=\{S_1,S_2,\cdots \}$, actor and critic parameters $\theta_{old}=\theta$ and $\phi$, replay storage, update interval $i_{update}$, and learning rate $\eta$. \\
        \textbf{while} $i \leq i_{end}$ \textbf{do} \\ 
        %\quad Sample $S_i=(M_i,X_i,\alpha_i)$ from $\mathcal{D}$ and $p(\alpha)$ \\
        %\quad Pick $S_i$ from $\mathcal{D}$ \\
        \quad Sample $\alpha$ from $p(\alpha)$ \\
        \quad Compute $\hat{Z}$ with Eqs. (\ref{eq:ggnn_encoder}), (\ref{eq:concatenation}) and (\ref{eq:concatenation_alpha}) given $\hat{S}_i=(S_i,\alpha)$ \\
        \quad Compute action probabilities $\pi_{\theta_{old}}(A|\hat{S})$ given $\hat{Z}$ and sample an action $A$, \\
        %\quad Sample an action $A \sim \pi_{\theta_{old}}(A|\hat{S})$ \\
        \quad Execute the action $A$, and receive reward $R(\alpha)$ \\
        \quad Store $(\hat{S},A,R(\alpha))$ to the replay storage \\
        \quad \textbf{if} $i_{update}$ \textbf{then} \\
        %\qquad Compute objectives $\mathcal{L}_{act}$ and $\mathcal{L}_{crt}$ with Eqs. (\ref{eq:obj_loss_actor} and \ref{eq:obj_loss_critic}) given the stored data in the replay storage  \\
        \qquad Compute PPO objectives of actor $\mathcal{L}_{act}$ and critic $\mathcal{L}_{crt}$.\\
        \qquad Update $\theta=\theta-\eta\frac{\partial}{\partial \theta}(\mathcal{L}_{act})$, $\phi=\phi-\eta\frac{\partial}{\partial \phi}(\mathcal{L}_{crt})$, and $\theta_{old}=\theta$
    \end{algorithmic}
\end{algorithm}

\subsubsection{Training Algorithm}
The whole training algorithm of DP-AS agent is described in Algorithm \ref{alg:training_algorithm}. We follow the PPO training algorithm, except for the additional sampling of $\alpha$ and incorporation of the original state and preference, $\hat{S}=(S,\alpha)$.

\subsection{Dynamic Preference Power Management Agent}
\label{subsec:power_management_agent}
Beyond the DP-AS task, which has a 1-dimensional preference space, we apply our proposed method to the PM sub-module that has a 2-dimensional preference space. It controls the power consumption of the network in addition to the SLA and resource usage (Section \ref{subsubsec:power_management_task}), so the reward could be formulated as follows:
\begin{equation}\label{eq:power_management_reward}
    R = -\frac{1}{|\mathcal{Q}|}\sum_{q \in \mathcal{Q}}{\frac{\zeta({p_q})}{\zeta_{SLA}}}-\alpha\sum_{f \in \mathcal{F}}{N(f)}-\beta\sum_{n \in \mathcal{N}}W(n),
\end{equation}
where $W(n)$ is an estimation function of power consumption at the node $n$. This power consumption could be estimated by the CPU load level and power consumption specification that is used in  \cite{abbas2022autonomous}, when the agent is trained based on simulation dataset. $\beta$ is the preference to control the importance of the power penalty. 

Except for the reward formulation, additional preference $\beta$, and its distribution $p(\beta)$, we follow the same utilization of $\alpha$ in DP-AS sub-module, such as policy and value networks and training algorithm. Considering the additional preference $\beta$ in addition to $\alpha$, even though there could be a correlation between them, we assume the independent relationship between those two for simplicity during the numerical process of estimating their distributions which we will explain in the next section.

\begin{figure}[h]
    \centering 
    \includegraphics[width=0.8\linewidth]{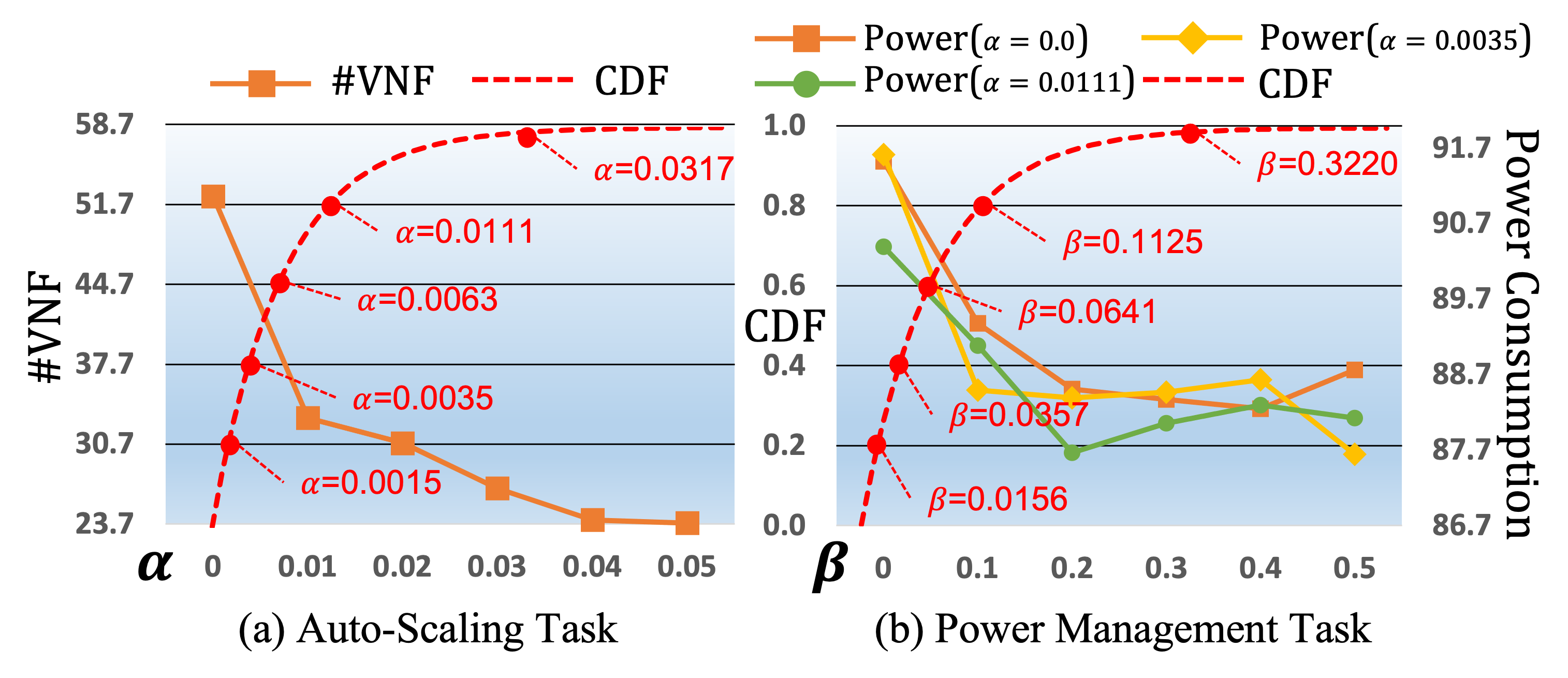}
    %\hbox{\small\hspace{0.1in} (a) Auto-scaling \hspace{0.7in} (b) Power management}
    \caption{Examples of collected data points (solid lines) and estimated exponential CDFs (dotted lines). %(Section \ref{subsec:parameter_distribution}). 
    These are examples of the Internet2 topology datasets. The red dots indicate that the estimated distribution's CDF values are 0.2, 0.4, 0.6, 0.8, and 0.99, respectively.}
    \label{fig:examples_of_collected_datapoints}
\end{figure}

\subsection{Numerical Estimation of Preference Distributions}
\label{subsec:preference_distribution}
To train an agent to be effective and generalizable for any preference during testing, it is important to design a proper preference distribution.
%To train an agent that is generalizable for any preference during testing, it is plausible to train the agent on the preference distribution, in which the preference's effect is uniformly distributed within a range.
Here, we hypothesize the `proper' preference distribution is the distribution that the effect of preference is uniformly distributed. Specifically, the effect means the actual outcome caused by an agent that is trained with the preference, such as the average resource usage at testing. If the effect is not uniformly distributed, the agent would be biased to a specific range of the effect. To estimate the preference distribution that transforms into the uniform effect distribution at the end of the training, we need to know the black-box dynamic between the preference and its effect. We propose a numerical method to approximate the black-box dynamic using several pre-experiment results as the observed samples from it. 

As a specific example of the AS task, we ran several experiments with fixed $\alpha=\left[0.0,0.01,0.02,0.03,0.04,0.05\right]$ (See Section \ref{sec:experiments_and_results} for the detail of experiments). After the training was done, we evaluated them and collected the total average number of VNF instances from the testing set. Let $V$ be the average VNF instance numbers after subtracting the minimum value as an offset. We view $V$ as one of the effects of preference in the network environment, and it is correlated with $\alpha$ because $\alpha$ is directly multiplied by the number of VNF instances in the reward as in Eq. (\ref{eq:auto_scaling_reward}). 

Fig. \ref{fig:examples_of_collected_datapoints}(a) demonstrates the observed data samples (solid line). We set the support of $V$ to $\left[0.0,V_{max}\right]$ where $V_{max}=max(V_0,V_1,\cdots)$. Then, as mentioned above, we define $p(V)=Unif_{\left[0.0,V_{max}\right]}(V)=\frac{1}{V_{max}}$ for the uniformly distributed effect of preference. Based on the shape, we model the black-box dynamic as an exponential function that maps $\alpha$ to $V$ as follows:
\begin{equation}
    V = f(\alpha) = V_{max}e^{-\lambda \alpha}, \label{eq:exponential_function}
\end{equation}
where $\lambda$ controls the stiffness of the exponential function. From the definition $p(V)$, we can derive the model of $p(\alpha)$ as follows:
\begin{equation}
    %p(\alpha) &= -p(V)\frac{\partial f(\alpha)}{\partial \alpha} \nonumber \\
    %&= -\frac{1}{V_{max}}(-\lambda V_{max})e^{-\lambda \alpha} \nonumber \\
    %&= \lambda e^{-\lambda \alpha}.
    p(\alpha) = p(V) \left| \frac{\partial f(\alpha)}{\partial \alpha} \right|
    %&= \frac{1}{V_{max}}(\lambda V_{max})e^{-\lambda \alpha} \nonumber \\
    = \lambda e^{-\lambda \alpha}.
    \label{eq:exponential_distribution}
\end{equation}
As a result, $p(\alpha)$ has the exponential distribution form. We estimated the value of $\lambda$ by applying the gradient descent method to minimize the squared error loss of the exponential model's output (Eq. \ref{eq:exponential_function}) and the observed value $V$. Consequently, we obtained $\lambda_{\alpha}=145.45$ and $\lambda_{\alpha}=241.05$ for the training of Internet2 and MEC topology datasets, respectively. Because we assumed the independent relationship between $\alpha$ and $\beta$, we conducted the same process for the distribution of $\beta$ with the setting of $\lambda_{\alpha}=145.45$ and the effect of $\beta$ is represented by the average power consumption (the last term in Eq. \ref{eq:power_management_reward}). We observed pre-experiment results by setting $\beta=\left[0.0,0.1,0.2,0.3,0.4,0.5\right]$. Consequently, we obtained $\lambda_{\beta}=42.51$ for the training of Internet2 topology dataset. The dotted lines in Fig. \ref{fig:examples_of_collected_datapoints} indicate the estimated distribution's cumulative density function (CDF) for $\alpha$ and $\beta$, respectively. The values on the dotted lines indicate 0.2, 0.4, 0.6, 0.8, and 0.99 CDFs.

\section{Experiments and Results}
\label{sec:experiments_and_results}
We test our proposed ideas on the two network management sub-modules: AS and PM. %We use network traffic simulation datasets as the environment of these tasks. 
As our goal is to train the agent to be generalizable on various preferences, we evaluate our trained agents on several tests with different preference settings. To understand the advantage of our proposed method, we compare our agents with the previous static preference RL algorithms \cite{seo2022updating} as baselines. 

\subsection{Simulation Data Descriptions}
\label{subsec:data_description}
To generate simulation datasets, we follow the previous works \cite{heo2020graph,heo2020reinforcement}. Based on each network topology as shown in Fig. \ref{fig:network_topologies}, we apply the public network traffic pattern collection\footnote{www.cs.utexas.edu/$\sim$yzhang/research/AbileneTM/} to resemble real-world service request pattern over time. 
Each active network traffic at each time step is assigned with a random service request consists of randomly selected source and destination nodes, a random bandwidth, and a random choice among a pre-defined sequence of requested VNF types. Note that SLA is usually determined by the top 5\% of total traffic delays from the generated dataset, where the longer delays than the other 95\% are considered as violated cases \cite{lee2021deep}. The set of VNF types $\mathcal{F}$ (middlebox) consists of \textit{firewall}, \textit{intrusion detection system} (IDS), \textit{proxy}, \textit{network address translation} (NAT), and \textit{wide area network optimization} (WANO). Based on the VNF types, we designed the four service types: (1) NAT-firewall-IDS, (2) NAT-proxy, (3) NAT-WANO, and (4) NAT-firewall-WANO-IDS.

After optimizing the VNF deployments with a CPLEX solver (dynamic programming method) \cite{bari2015ilp}, we perturb the deployments to be sub-optimal by randomly scaling the VNF instances as in \cite{seo2022updating}. 
Note that it is hard to use the dynamic programming method in a real-world scenario every time because of its high computational complexity. Additionally, in our dynamic preference scenario, the preference of each sub-reward factor varies; an optimal VNF deployment may not be optimal anymore if the preference changes. Finally, we have 14,736 data samples for both Internet2 topology and MEC topology datasets, respectively. We split them with an 8:1:1 ratio for training/validation/testing datasets. We trained and tested agents on the Internet2 and MEC datasets individually. % for the AS task. For the PM task, we used the Internet2 dataset.

\subsection{Agents and Hyperparameters}
For the hyperparameters of all AS agents, including the baseline (static preference) and our DP-AS, we used the same hyperparameters as in the previous work \cite{seo2022updating}, except the static values of $\alpha=\{0.0, 0.0015, 0.0035, 0.0063,$ $ 0.0111, 0.0317\}$ used for the baseline. These specific static values were selected based on the pre-experiments done in Section \ref{subsec:preference_distribution}. For our DP-AS agent, we set two distributions for $\alpha$, the estimated exponential (Section \ref{subsec:preference_distribution}) and uniform ($Unif_{\left[0.0,0.05\right]}(V)=\frac{1}{0.05}$. The uniform distribution setting is for the comparison between the estimated exponential distribution to show how our numerical method is advantageous. About our proposed power-management sub-module, since it is strongly based on the AS framework, we used the same hyperparameters, except the static values of $\beta=\{0.0, 0.0009, 0.0021, 0.0038, 0.0067, 0.0191\}$ based on the pre-experiment results. Similar to the DP-AS agent, for $\beta$, we used the estimated exponential distribution (Section \ref{subsec:preference_distribution}) and an arbitrary uniform distribution, $Unif_{\left[0.0,0.05\right]}(V)=\frac{1}{0.05}$, for our DP-PM agent's training. We ask readers to refer to the previous work \cite{seo2022updating} for the details of the hyperparameters.

\begin{figure}[h]
    \centering 
    \includegraphics[width=0.65\linewidth]{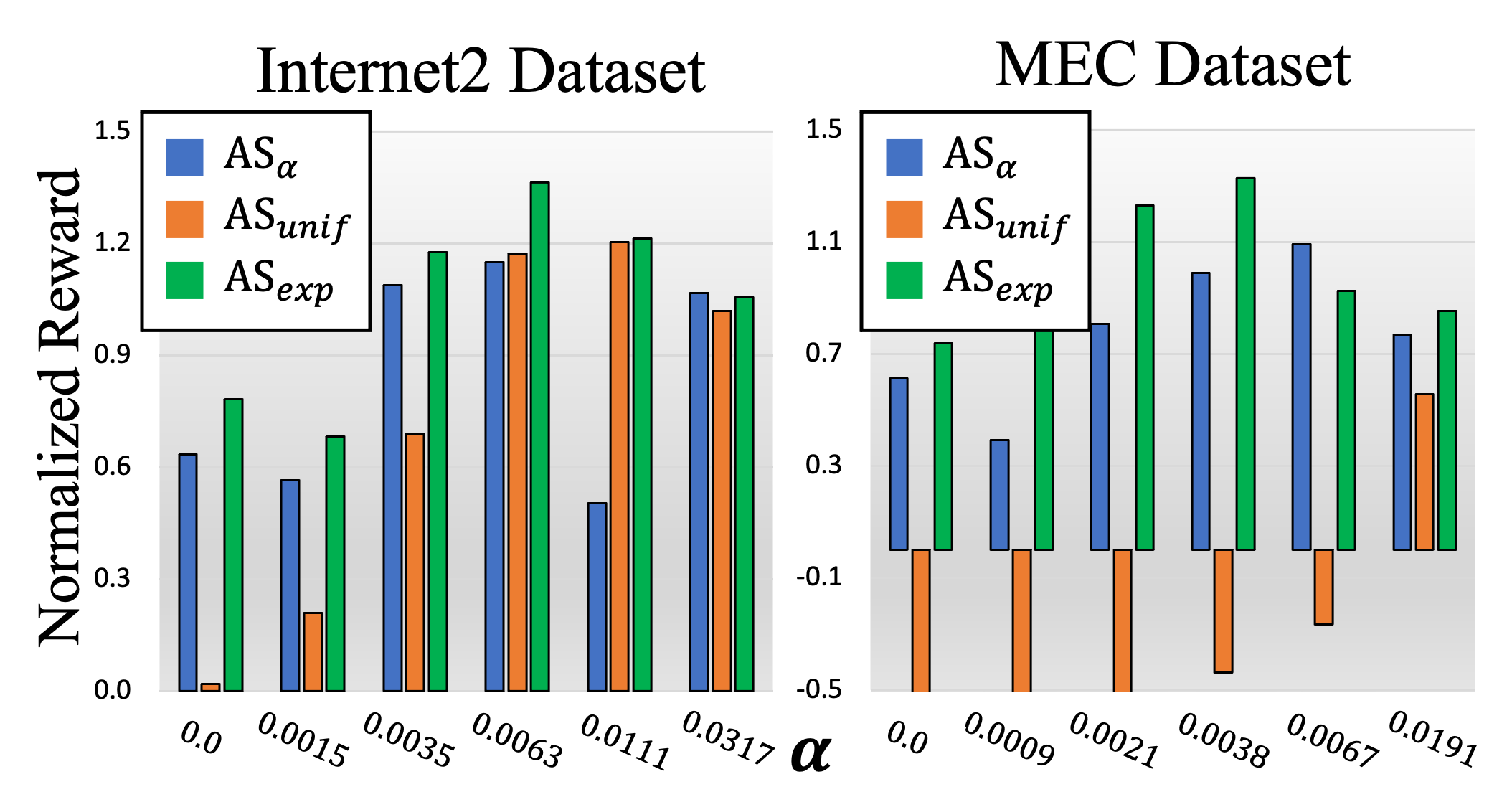}
    %\hbox{\small\hspace{0.1in} (a) Internet2 Dataset \hspace{0.7in} (b) MEC Dataset}
    \caption{Testings with static preference settings on the AS task. The $\alpha$ value of the horizontal axis indicates the preference used for that testing. $AS_{\alpha}$ means the baseline agent that is trained with the same preference of that testing. $AS_{exp}$ and $AS_{unif}$ mean our DP-AS agent trained with the estimated exponential and non-informative uniform distributions, respectively.}
    \label{fig:static_AS_testings}
\end{figure}

As we mentioned in Section \ref{subsubsec:autoscaling_task}, the SFC path according to each service request is generated by an external SFC model. Following the previous work, we employed the graph neural network-based SFC model \cite{heo2020graph} and pre-trained it with the generated simulation dataset. Based on the 5\% SLA violation rule, we took the delay value that assigns the longest 5\% of total delays of generated SFC paths to the violation cases. Consequently, we set $\zeta_{SLA}$ to 1,125 and 51 milliseconds for the Internet2 and MEC topology datasets, respectively.

\subsection{Results and Analysis}
\label{subsec:experimental_results}

\subsubsection{Testings with Static Preferences}
We evaluated all agents on several tests that have static preferences, that is the reward formulation is fixed according to the value of preference. Since the baseline agent has no function of adjusting its action according to different preferences, we only evaluate the baseline agent that is trained with the same preference setting of the testing, which is an ideal but non-realistic situation. However, we evaluate our DP-AS and DP-PM agents in every setting of static preferences.

Fig. \ref{fig:static_AS_testings} summarizes the experiment results of the AS task in terms of normalized reward (transformed rewards to have the mean of 0 and the standard deviation of 1 based on all reward results evaluated on the same fixed setting). As shown in the graph, our proposed DP-AS agent, $AS_{exp}$, achieved similar or better rewards compared to the baseline agents, $AS_{\alpha}$, for all the static preference settings. This indicates that our DP-AS agent successfully adapts its action to various preferences. About the outperformed performances, we interpret that the proposed method gives the agent more chances to learn about diverse states than the baseline method, leading to better generalization. However, our DP-AS agent with the uniform preference distribution, $AS_{unif}$, looks biased to the range of high values of $\alpha$ and performs poorly on the range of small values. 

\begin{figure}[h]
    \centering 
    \includegraphics[width=0.9\linewidth]{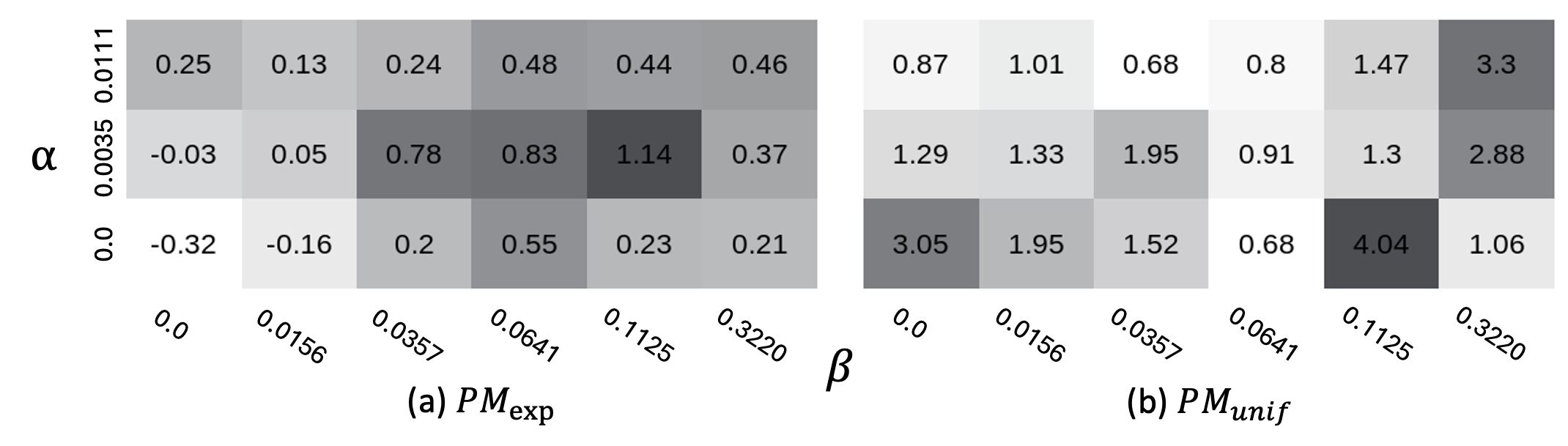}
    %\hbox{\small\hspace{0.25in} (a) $PM_{exp}$ \hspace{1.0in} (b) $PM_{unif}$}
    \caption{Normalized reward differences of the testings with static preference settings of the PM task on the Internet2 dataset. 
    %The vertical and horizontal axes indicate the value of $\alpha$ and $\beta$ used for that testing.
    %Instead of showing the values of normalized rewards, we show the difference in normalized rewards between our DP-PM agent and baseline agents. 
    The positive numbers mean that our DP-PM agent is better than the baseline agent in that testing.}
    \label{fig:static_PM_testings}
\end{figure}

Fig. \ref{fig:static_PM_testings} summarizes the experiment results of the PM task. We reported the difference in normalized rewards between our DP-PM agent and other baseline agents on the 2-dimensional grid for visualization. Similar to the AS task, our proposed DP-PM agent, $PM_{exp}$, achieved similar or better results than the baseline agents for all the settings, while $PM_{unif}$ is worse than all the baselines.

\begin{figure}[h]
    \centering 
    \hspace{-0.1in}
    \includegraphics[width=1.0\linewidth]{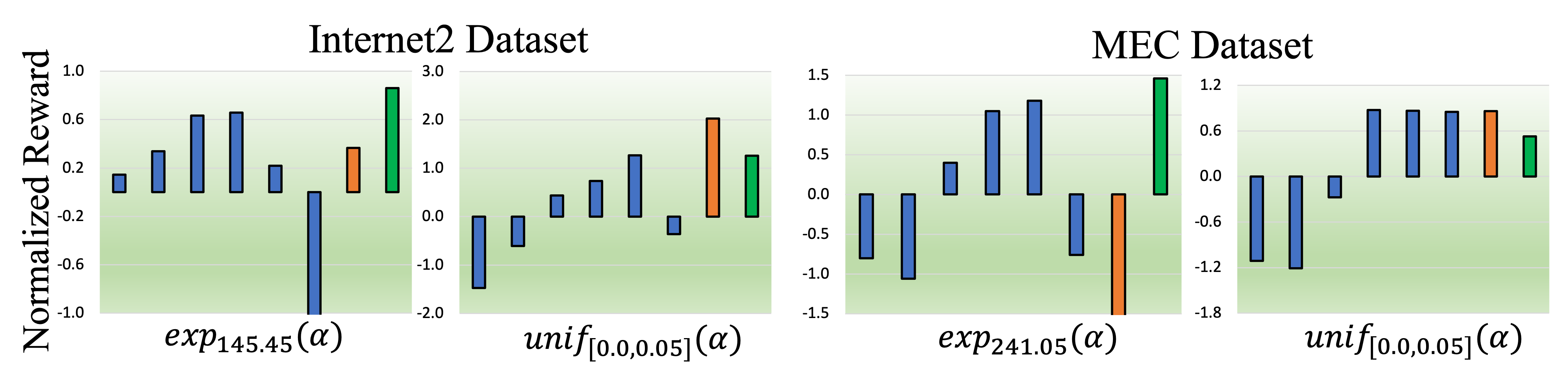}
    \caption{Testings with dynamic preference sampled from the exponential and uniform distributions (distributions written under each graph) on the AS task. Blue bars indicate baseline agents with different fixed $\alpha$ values. Refer the $\alpha$ values written in the horizontal axes of Fig. \ref{fig:static_AS_testings}. Orange and green bars indicate our DP-AS agents, $AS_{unif}$ and $AS_{exp}$, respectively.}
    \label{fig:dynamic_AS_testings}
\end{figure}

\subsubsection{Advantage of Providing a Better Distribution}
To better understand the advantage of the distribution estimation in Section \ref{subsec:preference_distribution}, we conducted additional tests with the dynamic preference setting where the preference is sampled from a distribution during testing. We set the two distributions we used to train our proposed DP-AS agent. Fig. \ref{fig:dynamic_AS_testings} demonstrates the results of the AS task in normalized reward. As expected, the DP-AS agents, $AS_{exp}$ and $AS_{unif}$, achieved their best score when it is tested on the same preference distribution setting as its training. However, while $AS_{exp}$ just slightly degrades the performance on the $Unif_{\left[0.0,0.05\right]}$ setting, $AS_{unif}$ significantly degrades on the $exp_{241.05}$ setting. Combined with the results, we understand that providing a better preference distribution can achieve better generalizability, and our proposed numerical method is a good candidate for estimating a good preference distribution compared to a non-informative distribution.

\begin{figure}[h]
    \centering 
    \includegraphics[width=0.75\linewidth]{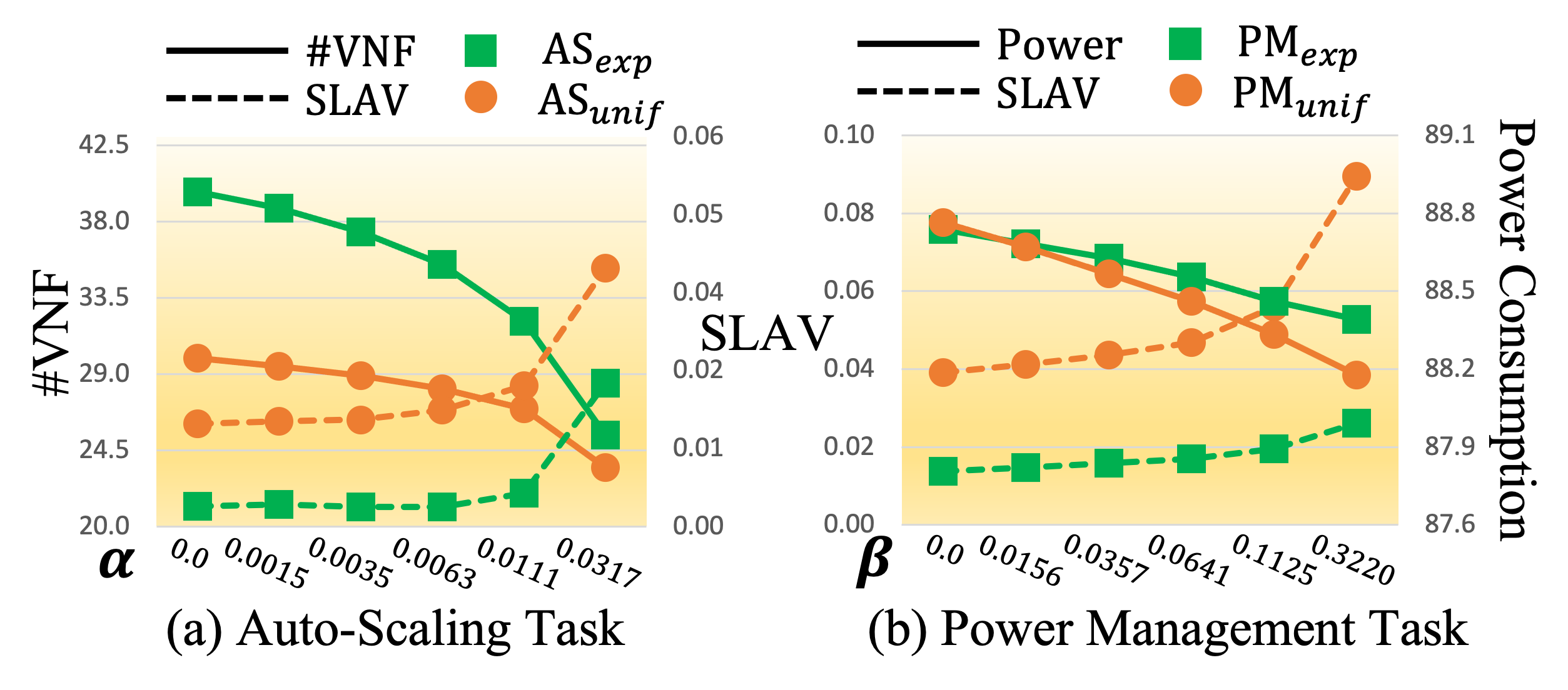}
    %\hbox{\small\hspace{0.1in} (a) Auto-Scaling  \hspace{0.7in} (b) Power management}
    \caption{DP agents' output values of sub-reward terms, the number of VNF instances, SLA violence rate, and power consumption.}
    \label{fig:other_factors_on_static_testings}
\end{figure}

\subsubsection{Analyses of Network Management Objectives}
In addition to the analysis of rewards, we analyze how our DP-AS and DP-PM agents change their actions according to different preference settings. Fig. \ref{fig:other_factors_on_static_testings} displays how $AS_{exp}$, $AS_{unif}$, $PM_{exp}$, and $PM_{unif}$ control the output factors of the network management environment, such as SLA violation rate (SLAV, rate of violated SFC paths over total paths), the number of VNF instances, and power consumption. As we expected, our dynamic preference agents dynamically react properly to the preference and result in different output factors. For example, when the testing $\alpha$ is high, $AS_{exp}$ decreases $\#VNF$ while increasing the SLAV. On the contrary, when the $\alpha$ is small, it increases $\#VNF$ and minimizes SLAV. A similar trade-off relationship between SLAV and power consumption is observed in the PM results.

\begin{figure*}[t]
    \centering 
    \includegraphics[width=0.8\linewidth]{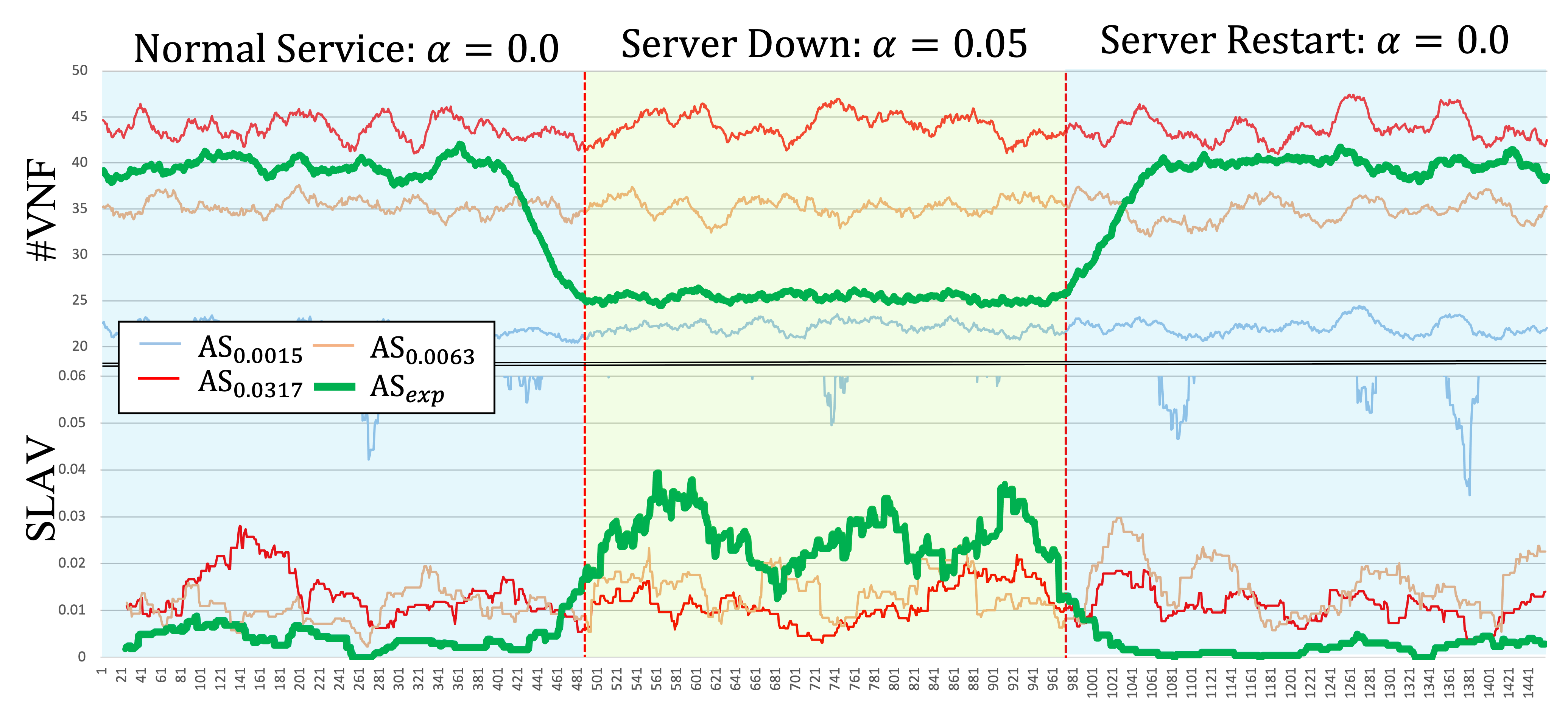}
    \caption{Testing while setting $\alpha$ manually based on the virtual scenario of server breaking down and restarting.
    %The graph of virtual scenario that a server breaks down and restarts.
    %Trajectory of testings with manual $\alpha$ manipulation of the AS task when a server breaks down and restarts. $AS_{exp}$ saves resources when the server is down and resumes the original behavior after the server restarts, while the other baseline agents do not change their behaviors for different situations.
    }
    \label{fig:manual_auto_scaling}
\end{figure*} 

In addition to the quantitative analysis above, we conducted a qualitative analysis to see the dynamics of our DP-AS agent based on a virtual scenario. In network management, a server that has deployed multiple VNF instances could break down due to accidents (e.g. blackout, hardware malfunction, and network attack). In such cases, it is suitable to save total resources rather than putting overloads on other running servers until the broken server is repaired and restarted. We designed this virtual scenario and manually defined the value of $\alpha$ to reflect it. Fig. \ref{fig:manual_auto_scaling} illustrates the trajectory of network management in this virtual scenario. $AS_{exp}$ agent behaves properly satisfying our original expectation, that is the agent saves resources when the server is down and resumes the original behavior after the server restarts, while the other baseline agents do not change their behaviors.

\section{Conclusion}
Unlike conventional reinforcement learning-based (RL) approaches of network management agents where the reward formulation is static with a fixed importance for each term, we include the possibility of dynamically changing importance factors. To make the network management agent adjustable to the dynamic setting of importance factors, we train the agent by inputting the value of importance to the policy model considering the importance as a variable. We also proposed the numerical method to estimate the importance distribution, and we use this estimated distribution during training to sample the value of importance from it. Our experiment results support our argument of adjustability inherent in our proposed approach with showing consistent outperformances across diverse preference settings.

\section*{Acknowledgment}
This research was supported by Basic Science Research Program through the National Research Foundation of Korea funded by the Ministry of Education (NRF-2022R1A2C1012633)

\bibliographystyle{splncs04}
\bibliography{references}

\end{document}